\documentclass[lettersize,journal]{IEEEtran}
\usepackage[latin9]{inputenc}
\usepackage{multirow}
\usepackage{amsmath}
\usepackage{amssymb}
\usepackage{graphicx}

\makeatletter

\providecommand{\tabularnewline}{\\}

\let\oldforeign@language\foreign@language
\DeclareRobustCommand{\foreign@language}[1]{%
  \lowercase{\oldforeign@language{#1}}}

\usepackage{xcolor}
\usepackage{multirow}
\usepackage[none]{hyphenat}
\usepackage{float}
\usepackage{dblfloatfix}
\usepackage{cite}
\usepackage{svg}
\usepackage{flushend}
\usepackage{textcomp}
\usepackage{t1enc}
\usepackage{times}

\PassOptionsToPackage{normalem}{ulem}
\usepackage{ulem}
    \providecolor{lyxadded}{rgb}{0,0,1}
    \providecolor{lyxdeleted}{rgb}{1,1,1}
    
    \DeclareRobustCommand{\lyxdeleted}[3]{{\color{lyxdeleted}\sout{#3}}}
\renewcommand{\lyxdeleted}[3]{}

\makeatother

\begin{document}
\raggedbottom 

\title{0.3-to-1.5-GHz LNA with Wideband Noise and Power Matching for Radio
Astronomy}
\author{Mark Lai, Vincent MacKay, Dallas Wulf, Peter Shmerko, Leonid Belostotski,
\emph{Senior Member, IEEE}\thanks{Manuscript received Feb. 9, 2023; revised Apr. 4, 2023; accepted Apr. 27, 2023. This work was funded in part by the University of Calgary, NSERC, and CMC Microsystems.}
\thanks{M.Lai, P. Shmerko, L. Belostotski are with the Department of Electrical and Software Engineering, University of Calgary, Calgary, AB, Canada (e-mail:(mark.lai,peter.shmerko, lbelosto)@ucalgary.ca). V. MacKay is with Department of Physics, University of Toronto (e-mail: vincent.mackay@mail.utoronto.ca). D. Wulf is with Department of Physics, McGill University (e-mail: dallas.wulf@mcgill.ca)}
\thanks{Color versions of one or more of the figures in this paper are available online at http://ieeexplore.ieee.org}}

\maketitle

\markboth{IEEE Microwave and Wireless Technology Letters,\  VOL. X, NO. XX,
XXX 202X}{Lai et al., \MakeLowercase{\textit{et al.}: \MakeUppercase{0.3-to-1.5-GHz
LNA with Wideband Noise and Power Matching for Radio Astronomy}}}



\begin{abstract}
This paper presents a low-noise amplifier (LNA) that was developed
for a new radio telescope comprised of 512 parabolic dish antennas.
The LNA closely interfaces to a custom-made antenna feed with an impedance
co-designed to provide noise matching over a 5:1 bandwidth. Additionally,
a method of broadband noise and power matching that allows the input
impedance to be controlled independently from the optimum signal-source
impedance to achieve minimum noise is also discussed. When measured
in a 50-$\Omega$ system, the LNA exhibits a return loss (RL) of \textgreater\,8\,dB
between 0.32 to 1.5\,GHz, S21 of 32\,dB\,$\boldsymbol{\pm}$\,1.2\,dB,
IP1dB \textgreater{} $\boldsymbol{-}$37\,dBm, and IIP3 \textgreater{}
$\boldsymbol{-}$20\,dBm. Noise parameter measurements show $\boldsymbol{T_{\text{min}}}$\,$\boldsymbol{\approx}$\,13$\boldsymbol{\pm}$4\,K
and noise temperatures $\boldsymbol{T_{50\Omega}}$\,$\approx$\,18$\pm$6\,K
between 0.5 to 1.4\,GHz.
\end{abstract}


\thispagestyle{empty} 
\global\long\def\contentsname{Contents}%
 
\global\long\def\listfigurename{List of Figures}%
 
\global\long\def\listtablename{List of Tables}%
 
\global\long\def\refname{References}%
 
\global\long\def\indexname{Index}%
 
\global\long\def\figurename{Fig.}%
 
\global\long\def\tablename{TABLE}%
 
\global\long\def\partname{Part}%
 
\global\long\def\appendixname{Appendix}%
 
\global\long\def\abstractname{Abstract}%
 \setlength{\parskip}{0pt} \bstctlcite{IEEEexample:BSTcontrol}

\section{Introduction}

The Canadian Hydrogen Intensity Mapping Experiment (CHIME) has revolutionized
the field of fast-radio-burst (FRB) detection. Following this success,
work began on the development of a next-generation telescope, the
Canadian Hydrogen Observatory and Radio-transient Detector (CHORD)
\cite{Vanderlinde2019mod}. The CHORD requires greater frequency coverage
and higher sensitivity. Thus, this paper presents a low-noise amplifier
(LNA) that satisfies the bandwidth (BW) and sensitivity requirements
of the CHORD and that can be installed on a feed antenna in a radome
and operated at outdoor temperatures. This LNA is designed to operate
between 0.3 to 1.5\,GHz and provide wideband noise and power matching
to the telescope antenna feed, which was co-designed alongside the
LNA to enable better noise matching.

The LNA described in this work operates over a 5:1 BW (i.e., a 5:1
ratio of upper-to-lower operating frequencies). While noise-cancelling
LNAs can operate over a 12:1 BW \cite{Liu2021}, the noise contribution
of their noise-cancelling circuit makes them unsuitable for radio
astronomy. Furthermore, there are several UWB (ultra-wide-band) LNA
topologies that operate over 3.4:1 BWs \cite{Bevilacqua2004,Ismail2004,JohnLong2006,2007Liao},
but they either cannot be noise matched over the whole BW or incur
losses and raise the LNA $T_{\text{min}}^{\text{LNA}}$ (i.e., minimum
noise temperature) above the transistor $T_{\text{min}}$. To reduce
the number of lossy components, feedback via intrinsic gate-drain
capacitance ($C_{\text{gd}}$) is employed in radio-astronomy LNAs
with 2:1 BWs \cite{Hu2006new,2022Sheldonmod,ThisaraKulatunga2018,BelostotskiANTEM}
to manipulate the LNA input impedance for broadband power matching,
while noise matching is implemented at the upper band edge to attain
a nearly flat noise figure (NF) in band.

\begin{figure}
\centering{}\includegraphics[bb=0bp 0bp 1064bp 354bp,width=0.95\columnwidth]{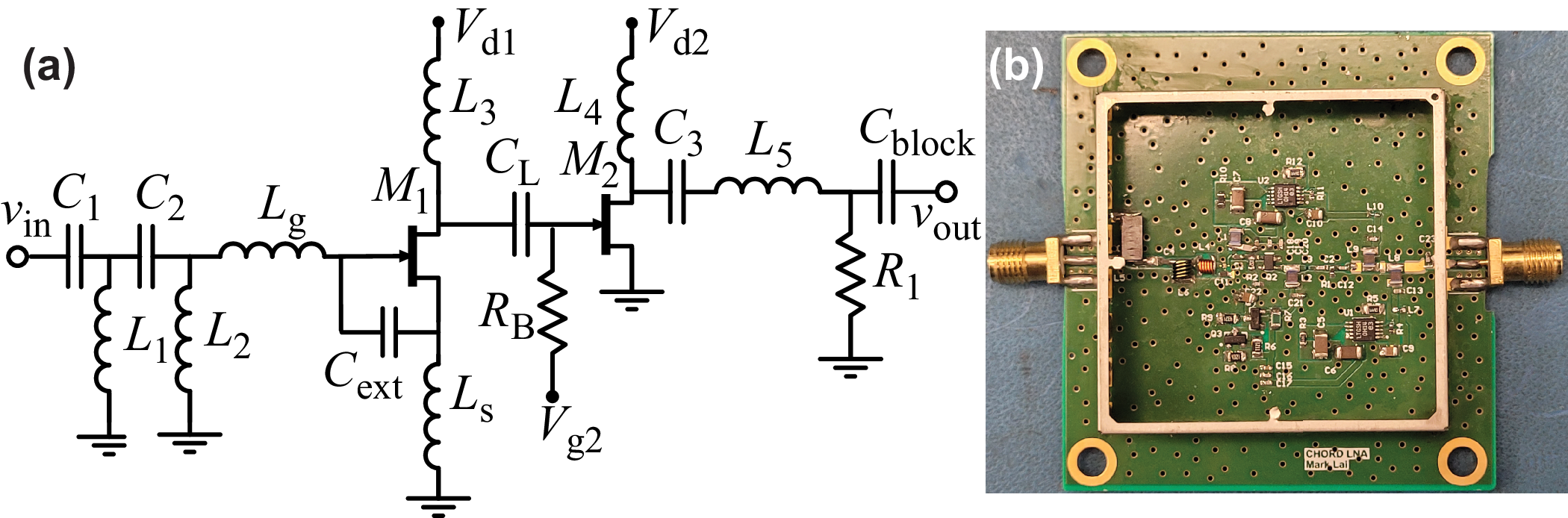}\caption{\label{completecircuit}(a) LNA circuit schematic (excluding biasing).
(b) Assembled LNA with its shield lid removed.}
\end{figure}

In the proposed LNA {[}see Fig.~ \ref{completecircuit}(a){]}, a
5:1 BW is achieved via a new combination of techniques employed in
narrower-band LNAs. Series-feedback is used to close the gap between
noise and power matching \cite{Boglione1997} and provides our starting
point, despite its narrowband operation. Since controlling transistor
load can be helpful in achieving simultaneous noise and power matching
via feedback elements \cite{Boglione1996,Engberg1974}, the transistor
load was designed such that the transistor input impedance approximated
the conjugate of the optimum source impedance ($Z_{\text{opt}}$)
for minimum noise. For the two-stage LNA, the input impedance of each
stage is partially manipulated via the Miller effect. Next, an input-matching
network based on \cite{Zailer2020} is employed at the input to mimic
the $-C$ behaviour of $Z_{\text{opt}}$. Finally, the antenna impedance
is further adjusted for better noise matching over the desired BW.


\section{LNA Description}


\subsection{\label{Cgd-feedback}Input impedance}

\begin{figure}
\centering \includegraphics[width=0.5\columnwidth]{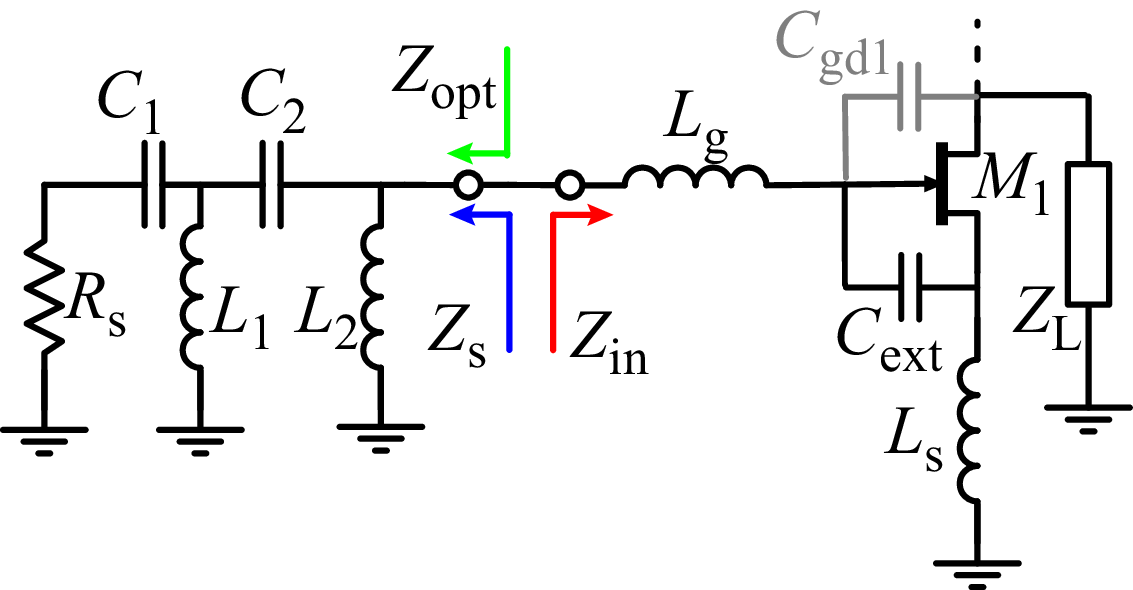}
\caption{\label{fig:matchingcircuit}Input stage of the LNA.}
\end{figure}

While the inductive-degenerated LNAs have been shown to enable simultaneous
noise and power matching, thereby coming close to achieving transistor
$T_{\text{min}}$, they achieve a narrowband power match due to the
impedances of the gate and source inductors ($L_{g}$ and $L_{s}$)
and the gate-source capacitance ($C_{\text{gs}}$) canceling out only
at their resonant frequency \cite{Belostotski2006TCAS}. However,
an additional analysis \cite{Hu2006new,Belostotski2006mod} revealed
another resonant mode at a lower frequency due to the Miller effect
caused by the intrinsic $C_{\text{gd}}$. Manipulating the load of
the LNA 1$^{\text{st}}$ stage causes at least two resonances appear
at the input. The 1$^{\text{st}}$ resonance---namely, that between
$C_{\text{gs}}$ and $L_{s}+L_{g}$---is noted in the conventional
description of the input impedance, while the 2$^{\text{nd}}$ resonance
appears due to the Miller effect of the 1$^{\text{st}}$ stage (i.e.,
$M_{1}$ and $L_{g}$, $L_{s}$, and $L_{3}$ in Fig. \ref{completecircuit}(a)).
Since the Miller effect is highly dependent on the 1$^{\text{st}}$-stage
load, multiple resonances in the input impedance can be formed throughout
the band without introducing loss at the LNA input, thereby improving
the LNA power match. In a recent work \cite{2022Sheldonmod}, the
input impedance ($Z_{\text{in}}$) {[}Fig. \ref{fig:matchingcircuit}{]},
seen when looking into $L_{\text{g}}$ and for an arbitrary load impedance
$Z_{L}$ on the output of the 1$^{\text{st}}$ stage, was derived
as 
\begin{equation}
Z_{\text{in}}\approx j\omega\left(L_{g}+L_{s}\right)+\frac{1}{j\omega C_{\text{gg}}}+\frac{g_{\text{m1}}L_{s}}{C_{\text{gg}}}\frac{r_{\text{o1}}}{Z_{L}}+\frac{r_{\text{o1}}}{j\omega C_{\text{gg}}Z_{L}}.\label{zinLNA}
\end{equation}
In \eqref{zinLNA}, $g_{\text{m1}}$ and $r_{\text{o1}}$ are the
transconductance and output resistance of $M_{1}$, $C_{\text{gg}}=C_{\text{gsT}}+(g_{\text{m1}}r_{\text{o1}}+1)C_{\text{gd1}}$,
$C_{\text{gsT}}=C_{\text{gs1}}+C_{\text{ext}}$ is the total gate-source
capacitance of $M_{1}$, and $Z_{L}$ is the impedance seen looking
into $C_{L}$ in Fig. \ref{completecircuit}(a). We make use of \eqref{zinLNA}
to realize the desired $Z_{\text{in}}$ by adjusting $Z_{L}$, which
depends on its own load via the Miller effect through, 
\begin{equation}
Z_{L}=\frac{1}{j\omega C_{L}}+\frac{1}{C_{\text{gs2}}}||\left[\frac{Z_{\text{out}}}{g_{\text{m2}}Z_{\text{out}}+1}+\frac{1}{j\omega C_{\text{gd2}}\left(g_{\text{m2}}Z_{\text{out}}+1\right)}\right].\label{eq:ZL}
\end{equation}
In \eqref{eq:ZL}, $L_{3}$ is assumed as a choke inductor, $g_{\text{m2}}$
and $C_{\text{gs2}}$ are the transconductance and gate-source capacitance
of $M_{2}$, and $Z_{\text{out}}=r_{\text{o2}}||\left(\frac{1}{j\omega C_{3}}+R_{\text{eq}}+j\omega L_{5}\right)$
represents the impedance seen at the drain of $M_{2}$ and $R_{\text{eq}}$
represents the parallel combination of $R_{1}$ and the termination
resistance at the output port. As the input impedances of 1$^{\text{st}}$
and 2$^{\text{nd}}$ stages depend on their loads, each stage cannot
be designed individually. Furthermore, as can be deduced through rather
involved expressions for $Z_{\text{in}}$, $Z_{L}$, and $Z_{\text{out}}$,
exact and concise solutions for most of the design parameters are
not available. Moreover, the selection of off-the-shelf (OTS) transistor
sizes is limited.

Nonetheless, it is possible to surmise some design guidelines. For
instance, the 2$^{\text{nd}}$ term in \eqref{eq:ZL} dominates and
has a real component. This term is the input impedance of the 2$^{\text{nd}}$
stage that is realized via the Miller effect on $C_{\text{gd2}}$
and can be manipulated by adjusting $C_{3}$ and $L_{5}$. In our
design, however, $C_{3}$ and $L_{5}$ are also used to flatten the
LNA gain at the upper band edge to counteract gain roll-off in the
first two stages. Therefore, the selection of $C_{3}$ and $L_{5}$
is constrained. The high-frequency resonance is mainly due to the
first three terms in \eqref{zinLNA}. At low frequencies, $Z_{L}$
is mainly capacitive due to $C_{L}$. A capacitive $Z_{L}$ creates
a real part in $Z_{\text{in}}$ due to the Miller effect {[}via the
4$^{\text{th}}$ term in \eqref{zinLNA}{]} and increases the total
inductance in $Z_{\text{in}}$ {[}via the 3$^{\text{rd}}$ term in
\eqref{zinLNA}{]}, thus enabling a low-frequency resonance. The circuit
exhibits another resonance near the midband where both terms in \eqref{eq:ZL}
are not negligible. These resonances help broaden the input matching
of the LNA and show that the input impedance $Z_{\text{in}}$ can
be manipulated by tuning $C_{L}$, $C_{3}$, and $L_{5}$. The further
tuning of $Z_{\text{in}}$ during simulations is enabled by $L_{g}$,
$L_{s}$, and $C_{\text{ext}}$, which was added to fine tune $C_{\text{gsT}}$.


\subsection{Simultaneous noise and power matching}

\begin{figure}
\begin{centering}
\includegraphics[width=0.6\columnwidth]{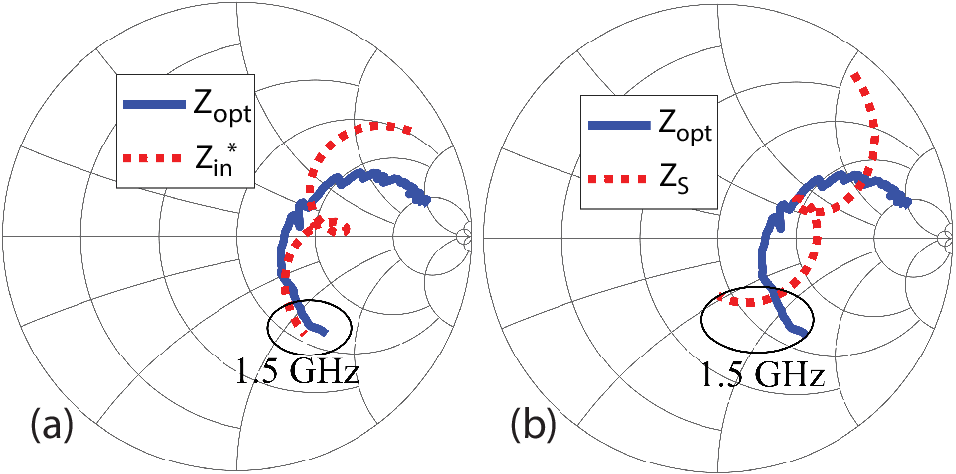}
\par\end{centering}
\caption{\label{fig:Sopt}Simulated (a) $Z_{\text{opt}}$ vs $Z_{\text{in}}^{*}$
and (b) $Z_{\text{opt}}$ vs $Z_{\text{s}}$.}
\end{figure}

Section \ref{Cgd-feedback} discussed the possibility of manipulating
the LNA $Z_{\text{in}}$ via $Z_{L}$. If the 1$^{\text{st}}$ stage
has a high gain, the manipulation of $Z_{L}$ will not affect the
LNA noise parameters, including $Z_{\text{opt}}$ (identified in Fig.
\ref{fig:matchingcircuit})\cite{Belostotski2006mod}. As a result
$Z_{\text{in}}$ can be tuned by manipulating $Z_{L}$ such that $Z_{\text{in}}^{*}\approx Z_{\text{opt}}$
for simultaneous noise and power matching. To minimize the LNA noise
temperature, $Z_{\text{opt}}$ and $Z_{\text{in}}$ are optimized
near the upper band edge where $T_{\text{min}}$ is the highest. As
shown in Fig. \ref{fig:Sopt}(a), $Z_{\text{opt}}$ and $Z_{\text{in}}^{*}$
approximate each other well at higher frequencies but deviate at the
lower band edge where transistor noise is lower and sky temperature
is higher, thereby making the LNA noise performance less critical.

Since tuning the OTS transistor for simultaneous noise and power matching
prevents the LNA from realizing $Z_{\text{in}}\approx$ 50\,$\Omega$,
a matching network is needed to transform $R_{s}=50\Omega$ to track
$Z_{\text{opt}}$, which exhibits a $-C$ behaviour. To approximate
the $-C$ behaviour, the matching network and $R_{S}$ form low-Q
networks \cite{Zailer2020}. Unlike \cite{Zailer2020}, which employed
a matching network comprised of packaging parasitics, our design,
which uses OTS transistors, employs surface-mounted components. Series-$C$
(i.e., $C_{1(2)}$) shunt-$L$ (i.e., $L_{1(2)}$) networks are used
for matching, as they provide both dc blocking and gate bias (e.g.,
via $L_{2}$) and are also less lossy compared to series-$L$ shunt-$C$
networks. Fig. \ref{fig:Sopt}(b) shows the resultant transformation
of $R_{s}=50\Omega$ to $Z_{s}$ (see Fig. \ref{fig:matchingcircuit}),
approximating the $-C$ behaviour of $Z_{\text{opt}}$ and reaching
$Z_{\text{opt}}$ at three frequencies within the band. While this
matching network introduces some additional insertion loss and noise,
the benefits gained from noise matching using this method (vs narrow-band
matching) is seen toward the band edges, where lower NFs are observed.



\section{\label{sec:results}Measurement Results}

\begin{figure}
\begin{centering}
\includegraphics[width=0.9\columnwidth]{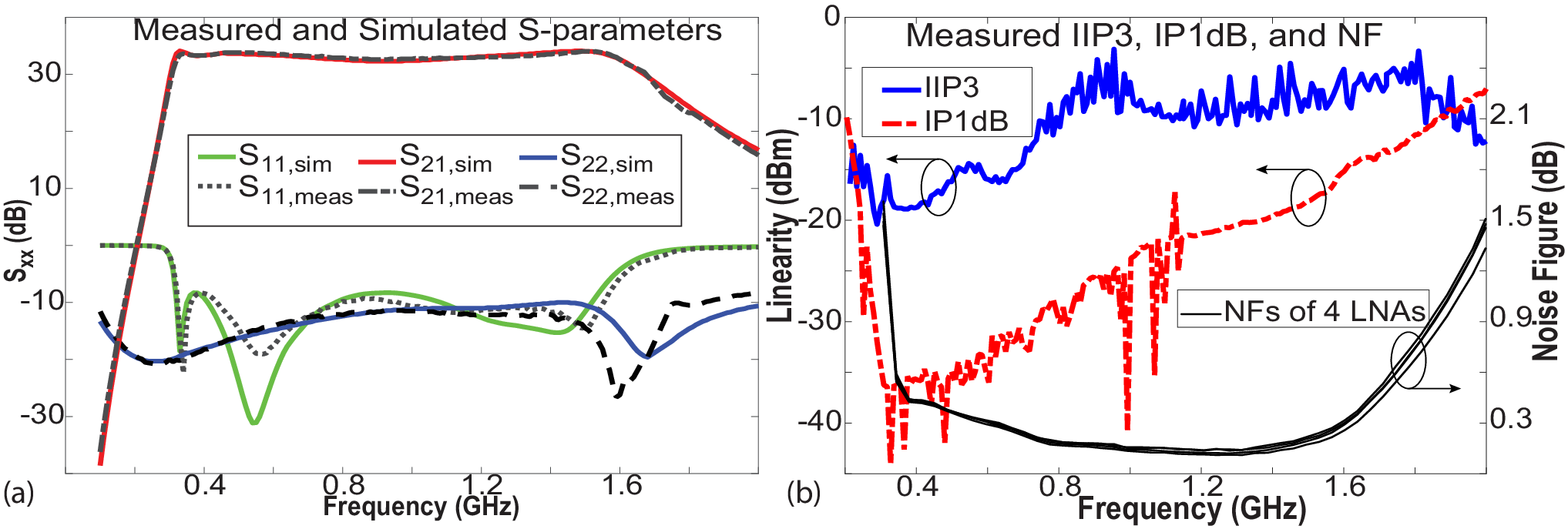}
\par\end{centering}
\caption{\label{fig:sparam}(a) S-parameters. (b) IP1dB, IIP3 at a 0.1~MHz
two-tone spacing, and NFs at the connectors.}
\end{figure}

As the CHORD requires 1500 LNAs, some design choices were constrained
by part costs. The LNA was assembled on a Rogers 4003C PCB {[}Fig.
\ref{completecircuit}(b){]} using Diramic's low-noise (2.3\,K\,$\leq$\,$T_{\text{min}}$\,$\leq$\,4\,K)
6-finger 0.4-mm wide pHEMT InP transistor for $M_{1}$ and a less-expensive
MiniCircuits SAV-541+ pHEMT transistor for $M_{2}$, which was selected
by conducting multiple simulations of numerous OTS transistors until
satisfactory design was achieved. Extensive EM simulations using non-ideal
component models were conducted to further optimize the design given
PCB and component parasitics. Onboard voltage regulation provides
the required biasing, and the whole assembly (with the regulators
and biasing circuitry) draws 90\,mA from a 5-V supply. The LNA core
consumes 255\,mW from a 4-V supply. High-Q capacitors and low-loss
air-core inductors were used in the matching network to avoid an increase
in noise.The measured S-parameters are plotted in Fig. \ref{fig:sparam}(a).
The LNA shows $S_{\text{21}}=33\pm1.2$\,dB within the band. When
measured in a 50-$\Omega$ system, an input RL\,\textgreater 8\,dB
from 0.32 to 1.5\,GHz and \textgreater 10\,dB was obtained in most
of the band. An output RL\,\textgreater\,12\,dB was also obtained
across the whole band. The group delay was \textless 2~ns from 0.4
to 1.5\,GHz but peaked at 6.5~ns below 0.4\,GHz. The linearity
metrics presented in Fig. \ref{fig:sparam}(b) show the LNA has an
IP1dB of at least -37\,dBm and an IIP3 greater than -20\,dBm. The
linearity is worse at lower frequencies due to the higher gain of
the 1$^{\text{st}}$ stage. The measured NFs are shown in Fig. \ref{fig:sparam}(b).

\begin{figure}
\centering\includegraphics[width=0.9\columnwidth]{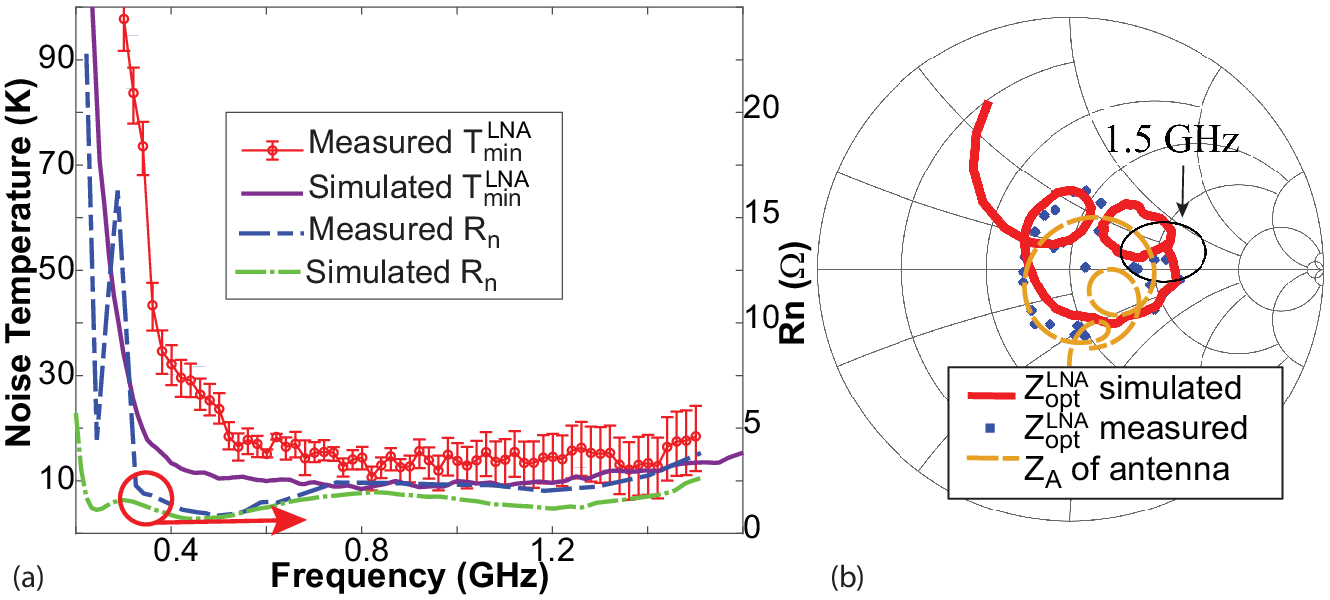}
\caption{\label{fig:meas-NP}(a) Average $T_{\text{min}}$ and $T_{50\Omega}$
and $R_{n}$. 6\,K error bars in are based on measurements with 3
noise sources and are dominated by the noise source uncertainty. (b)
LNA $Z_{\text{opt}}^{\text{LNA}}$ and antenna impedance $Z_{A}$.
The limited BW of the source-pull tuner adversely impacted accuracy
below 0.5\,GHz.}
\end{figure}

\begin{figure}
\begin{centering}
\includegraphics[clip,width=0.9\columnwidth]{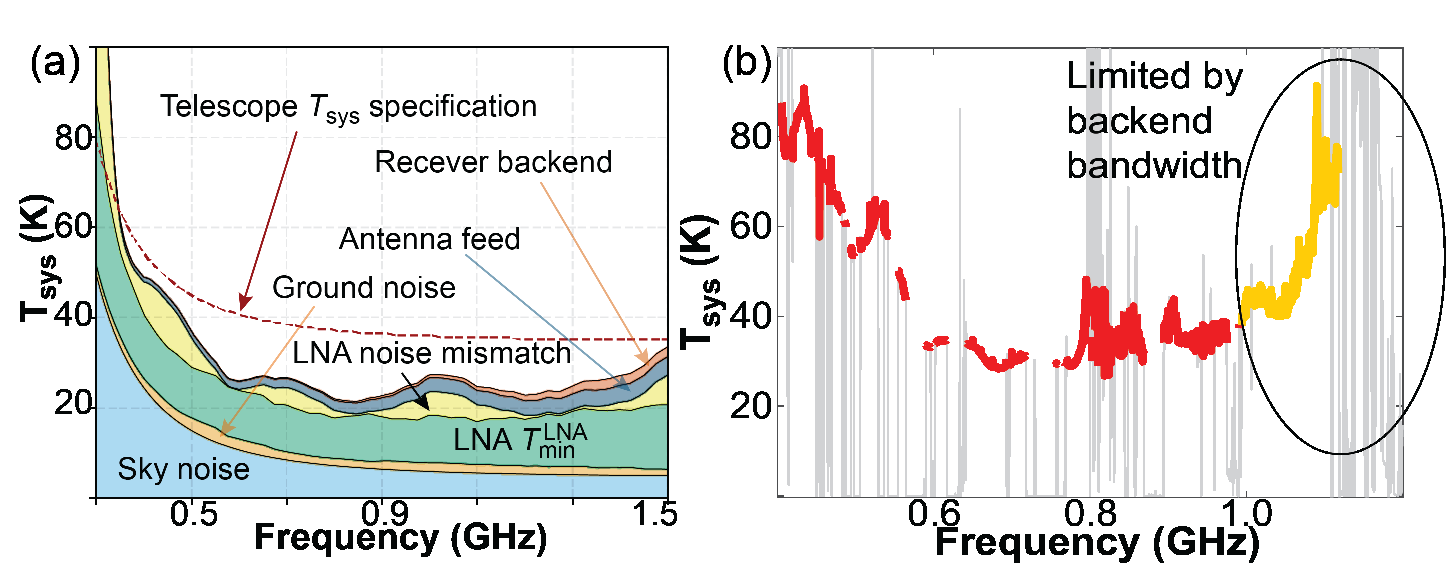}
\par\end{centering}
\caption{\label{fig:tsys}(a) $T_{\text{sys}}$ noise budget based on our LNA
and CHORD $T_{\text{sys}}$ specifications \cite{Mackay2022}. (b)
Measured $T_{\text{sys}}$ (preliminary results). Local RFI has been
greyed out for readability.}
\end{figure}

When assembled on the telescope, the antenna is soldered directly
to the LNA input. Thus, the input connector was de-embedded using
a custom-designed TRL kit to measure the LNA noise parameters ($T_{\text{min}}^{\text{LNA}}$,
equivalent noise resistance $R_{n}$, and $Z_{\text{opt}}^{\text{LNA}}$).
This measurement used a source-pull Y-factor method described in \cite{Belostotski2010TMTT}.
As seen in Fig. \ref{fig:meas-NP}(a), 
the LNA exhibits $T_{\text{min}}^{\text{LNA}}$ within (11 to 23)$\pm$6~K
between 0.5 and 1.5\,GHz. Three noise sources were used to conduct
these measurements and are responsible for the 6-K error bars. The
limited BW of the source-pull impedance tuner adversely impacted the
accuracy of measurements below 0.5\,GHz \cite{Himmelfarb2016a}.
Fig. \ref{fig:meas-NP}(b) shows the $Z_{\text{opt}}^{\text{LNA}}$
of the whole LNA, which illustrates the end result of the matching
network transforming $R_{s}$ to approximate the optimum source impedance
of the LNA core. In addition, the results in Fig. \ref{fig:meas-NP}(b)
also shows that $Z_{A}$ is realized by the telescope antenna. Based
on these results, the telescope noise budget is calculated and shown
in Fig. \ref{fig:tsys}(a) \cite{Mackay2022}. This budget includes
the LNA noise, losses in the antenna, the antenna's radiation efficiency,
noise from the ground at 300\,K illuminated by sidelobes and the
sky (modeled via an antenna solid beam \cite{Condon2016}, and noise
from the cosmic microwave background \cite{Fixsen_2009}, galactic
continuum \cite{Haslam1981}, and 300\,K atmospheric temperature
scaled by the zenith opacity \cite{Rosenkranz_1975}). 
$T_{\text{min}}^{\text{LNA}}$ and noise-mismatch penalty are shown
separately.

Two LNAs were used in the CHORD demonstrator, ``D3A6,'' which was
band-limited to 1.2\,GHz, to obtain a preliminary estimate of the
system noise temperature $T_{\text{sys}}$ in Fig. \ref{fig:tsys}(b).
The initial measurement was conducted using 3 dishes, with 2 LNAs
being used for each polarization on one dish and 4 noisier OTS LNAs
being used on the other 2 dishes. The measurements in Fig. \ref{fig:tsys}(b)
include contributions from the sky, the ground, feed losses, and local
interference. Furthermore, to extract the data, sidelobes were assumed
to make up 15\% of the solid angle, and the $x$ and $y$ polarizations
of the feed were assumed to be identical. These results demonstrate
that the proposed design is capable of achieving $T_{\text{sys}}\lesssim30$\,K.
The performances of the proposed LNA and other similar LNAs are summarized
in Table~\ref{tab:Summary-and-comparison}.

\begin{table}
\caption{\label{tab:Summary-and-comparison}Summary and comparison with other
room-temperature pHEMT LNAs for radio astronomy.}

\begin{centering}
\begin{tabular}{l||c|c|c|c}
\hline 
 & \multicolumn{1}{l|}{\textbf{\footnotesize{}This work}} & \multicolumn{1}{c|}{{\footnotesize{}\cite{Weinreb2021new}}} & \multicolumn{1}{c|}{{\footnotesize{}\cite{ThisaraKulatunga2018}}} & \multicolumn{1}{c}{{\footnotesize{}\cite{Wang2015}}}\tabularnewline
\hline 
\hline 
\multirow{1}{*}{{\footnotesize{}Tech.}} & \textbf{\footnotesize{}0.1$\mu$m InP} & {\footnotesize{}0.1$\mu$m InP} & {\footnotesize{}GaAs} & {\footnotesize{}0.15$\mu$m GaAs}\tabularnewline
\hline 
{\footnotesize{}Freq. (GHz)} & \textbf{\footnotesize{}0.3-1.5} & {\footnotesize{}0.8-1.6} & {\footnotesize{}0.4-0.8} & {\footnotesize{}1.6-3.1}\tabularnewline
\hline 
{\footnotesize{}BW} & \textbf{\footnotesize{}5:1} & {\footnotesize{}2:1} & {\footnotesize{}2:1} & {\footnotesize{}1.9:1}\tabularnewline
\hline 
{\footnotesize{}Max. Gain} & \textbf{\footnotesize{}34 dB} & {\footnotesize{}38 dB} & {\footnotesize{}41 dB} & {\footnotesize{}27.5 dB}\tabularnewline
\hline 
{\footnotesize{}$T_{50\Omega}$ (K)} & \textbf{\footnotesize{}(15 to 40)$\pm$6$^{+}$} & {\footnotesize{}7.5 to 13} & {\footnotesize{}18 to 23.5} & {\footnotesize{}48 to 63}\tabularnewline
\hline 
{\footnotesize{}Power} & \textbf{\footnotesize{}255 mW} & {\footnotesize{}255 mW} & {\footnotesize{}406 mW} & {\footnotesize{}51.8 mW}\tabularnewline
\hline 
{\footnotesize{}Form factor ($\text{mm}^{3}$)} & {\footnotesize{}67$\times$57$\times$11} & {\footnotesize{}86$\times$54$\times$22} & {\footnotesize{}72$\times$42$\times$5} & {\footnotesize{}N/A}\tabularnewline
\hline 
\multicolumn{5}{l}{{\footnotesize{}$^{+}$In a 50-$\Omega$ environment. The LNA is designed
to interface to the antenna,}}\tabularnewline
\multicolumn{5}{l}{{\footnotesize{}rather than 50-$\Omega$ equipment. Measurement errors
dominated by noise}}\tabularnewline
\multicolumn{5}{l}{{\footnotesize{}source uncertainty.}}\tabularnewline
\end{tabular}
\par\end{centering}
\end{table}


\section{Conclusion}

This paper presented a method that facilitates simultaneous broadband
power and noise matching. By controlling the load impedance of the
1$^{\text{st}}$ stage, the LNA input impedance can be manipulated
while $Z_{\text{opt}}$ remains constant until the input impedance
approaches $Z_{\text{opt}}^{*}$, thereby achieving simultaneous power
and noise matching. A matching network is then implemented to transform
the source impedance to approximate $Z_{\text{opt}}$ over frequency.
The results presented herein demonstrate that an LNA design based
on this method can achieve sub-20K noise temperatures between 0.5
to 1.4\,GHz, with an 8-dB RL between 0.32 and 1.5\,GHz.





\bibliographystyle{IEEEtran}
\bibliography{PaperBib}

\end{document}